# Effects of high-power laser irradiation on sub-superficial graphitic layers in single-crystal diamond

F. Picollo[1,2,3*], S. Rubanov[4], C. Tomba[5,6], A. Battiato[2,1,3], E. Enrico[7], A. Perrat-Mabilon[8,9], C. Peaucelle[8,9], T. N. Tran Thi [5,10], L. Boarino[4], E. Gheeraert[5,10], P. Olivero[2,1,3,7]

[1] *National Institute of Nuclear Physics (INFN), section of Torino, Italy*

[2] *Physics Department and "NIS" Inter-departmental centre, University of Torino, Torino, Italy*

[3] *National Interuniversity Consortium for the Physical Sciences of Matter (CNISM), Torino Unit, Italy*

[4] *Bio21 Institute, University of Melbourne, Australia*

[5] *University of Grenoble Alpes, F-38000 Grenoble, France*

[6] *Laboratoire des Technologies de la Microelectronique, Minatec Campus, F-38054 Grenoble, France*

[7] *National Institute of Metrologic Research (INRiM), Torino, Italy*

[8] *University of Lyon 1, CNRS, Inst Phys Nucl Lyon, F-69622 Villeurbanne, France*

[9] *IN2P3, F-69622 Villeurbanne, France*

[10] *CNRS, Institute NEEL, F-38042 Grenoble, France*

## Abstract

We report on the structural modifications induced by a λ = 532 nm ns-pulsed high-power laser on sub-superficial graphitic layers in single-crystal diamond realized by means of MeV ion implantation. A systematic characterization of the structures obtained under different laser irradiation conditions (power density, number of pulses) and subsequent thermal annealing was performed by different electron microscopy techniques. The main feature observed after laser irradiation is the thickening of the pre-existing graphitic layer. Cross-sectional SEM imaging was performed to directly measure the thickness of the modified layers, and subsequent selective etching of the buried layers was employed to both assess their graphitic nature and enhance the SEM imaging contrast. In particular, it was found that for optimal irradiation parameters the laser processing induces a six-fold increase the thickness of sub-superficial graphitic layers without inducing mechanical failures in the surrounding crystal. TEM microscopy and EELS spectroscopy allowed a detailed analysis of the internal structure of the laser-irradiated layers, highlighting the presence of different nano-graphitic and amorphous layers. The obtained results demonstrate the effectiveness and versatility of high-power laser irradiation for an accurate tuning of the geometrical and structural features of graphitic structures embedded in single-crystal diamond, and open new opportunities in diamond fabrication.

* corresponding author: picollo@to.infn.it

ph: +39 011 670 7879

fax: +39 011 670 7020

Keywords: diamond, high power laser, graphitisation, TEM microscopy

PACS codes: diamond, 81.05.ug

## 1. Introduction

Diamond is well known for its range of extreme mechanical, thermal and optical properties, which make it an attractive material for a variety of applications [1]. Nevertheless, diamond is a metastable allotropic form of carbon at standard pressure and temperature, and can be converted into graphite if an energy barrier is overcome [2]. Several approaches have been developed to induce this phase transition, among which ion-beam-induced graphitization [3–9] and laser-induced graphitization [10,11] play a prominent role. The former approach takes advantage of the ion-induced defect creation caused by nuclear collisions to amorphize the material and the subsequent thermal annealing to convert amorphized regions into a graphitic phase [3]. The latter approach is based on complex non-equilibrium dynamics induced by high-power light absorption, which were modelled with different theoretical approaches based on the non-radiative recombination of electron-hole pairs [11] or on a non-thermal ultrafast non-equilibrium phase transition [12,13].

Several previous studies explored the laser-induced graphitization process of single-crystal diamond: the first investigations dating back to the 80's were focused on realization of graphitic structures on diamond surface with direct writing or optical projection by means of excimer lasers [14], approached that was further investigated also in recent years [15]. Subsequently, new theoretical models of pulsed laser irradiation were proposed taking into account fast energy transfer mechanisms [16]. Consequently, in the last decade several works were carried out to exploit the possibility of realizing three-dimensional structures into diamond bulk by means of femtosecond [17,18] and picosecond [18–20] pulsed laser writing. Furthermore, the possibility of enhancing the resolution in the laser fabrication of graphitic structures with the use of adaptive optical elements was recently demonstrated [21,22].

Direct laser-induced graphitization represents an extremely versatile technique with promising applications in different fields such as the realization of diamond-based particle detectors [23–27] and (upon the selective removal of the graphite) microfluidics devices for biomedical sensing [28].

On the other hand, this technique is limited by the poor geometrical quality of structures finishing, which is inherently caused by the nature of the graphitization process [18,20]. In order to overcome this limitation, laser-induced graphitization in diamond can be combined with a preliminary MeV-ion-induced graphitization stage. By taking advantage of the high degree of control on the geometrical properties (depth, thickness) of MeV-ion-induced buried graphitic structures in diamond allowed by the peculiar nuclear energy loss profile of MeV ions [29], this double-step procedure guarantees a better definition in the material micro-structuring [30] and also represents an interesting improvement in the realization of particle detectors [26,31,32], bolometers [33,34], bio-sensors [35–37], metallic-dielectric structures [38] and microfluidics [39].

In the present paper we report on the use of ns-pulsed laser irradiation for the structural modification and thickening of sub-superficial graphitic layers in diamond, which were realized by means of MeV ion implantation. The above-mentioned structures are imaged before and after the selective removal of the graphitic phase with respect to the surrounding diamond matrix, and are characterized in their structural properties by transmission electron microscopy.

## 2. Experimental

In the present study, a commercial synthetic (001) single-crystal diamond grown by High Pressure High Temperature method (HPHT) by ElementSix (Ascot, UK) was used. The diamond is 3×3×0.3 $mm^3$ in size and it is classified as type Ib, having a nominal substitutional nitrogen concentration between 10 ppm and 100 ppm. The sample is cut along the 100 crystal direction and it is optically polished on the two opposite large faces.

The sample was implanted at room temperature at the "Service Faisceaux d'Ions" laboratory of the Nuclear Physics Institute (University of Claude Bernard Lyon 1) across one of the two main polished surfaces with a broad 2 MeV $He^+$ ion beam to deliver a uniform fluence of $1\times10^{17}$ $cm^{-2}$ across the irradiated area. During the implantation, the beam current was ~200 Na. The process of damage induced by MeV ions in matter occurs mainly at the end of ion range, where the cross section for nuclear collisions is strongly enhanced, after the ion energy is progressively reduced by electronic interactions occurring in the initial stages of the ion path [40]. Figure 1 shows the strongly non-uniform depth profile of the density of induced vacancies ($\#_{vac}$ $cm^{-3}$) evaluated in a linear approximation as the product between the implantation fluence ($\#_{ions}$ $cm^{-2}$) and the linear density of induced vacancies per single ion ($\#_{vac}$ $cm^{-1}$ $\#_{ions}^{-1}$). The latter quantity was estimated with the "Stopping and Range of Ions in Matter" (SRIM)-2013.00 Monte Carlo code [29] in "Detailed calculation with full damage cascade" mode by taking an atom displacement energy value of 50 Ev [41]. The high density of damage induced by ion implantation promotes the conversion of the diamond lattice to an amorphous phase, which is located ~3.5 µm below the sample surface.

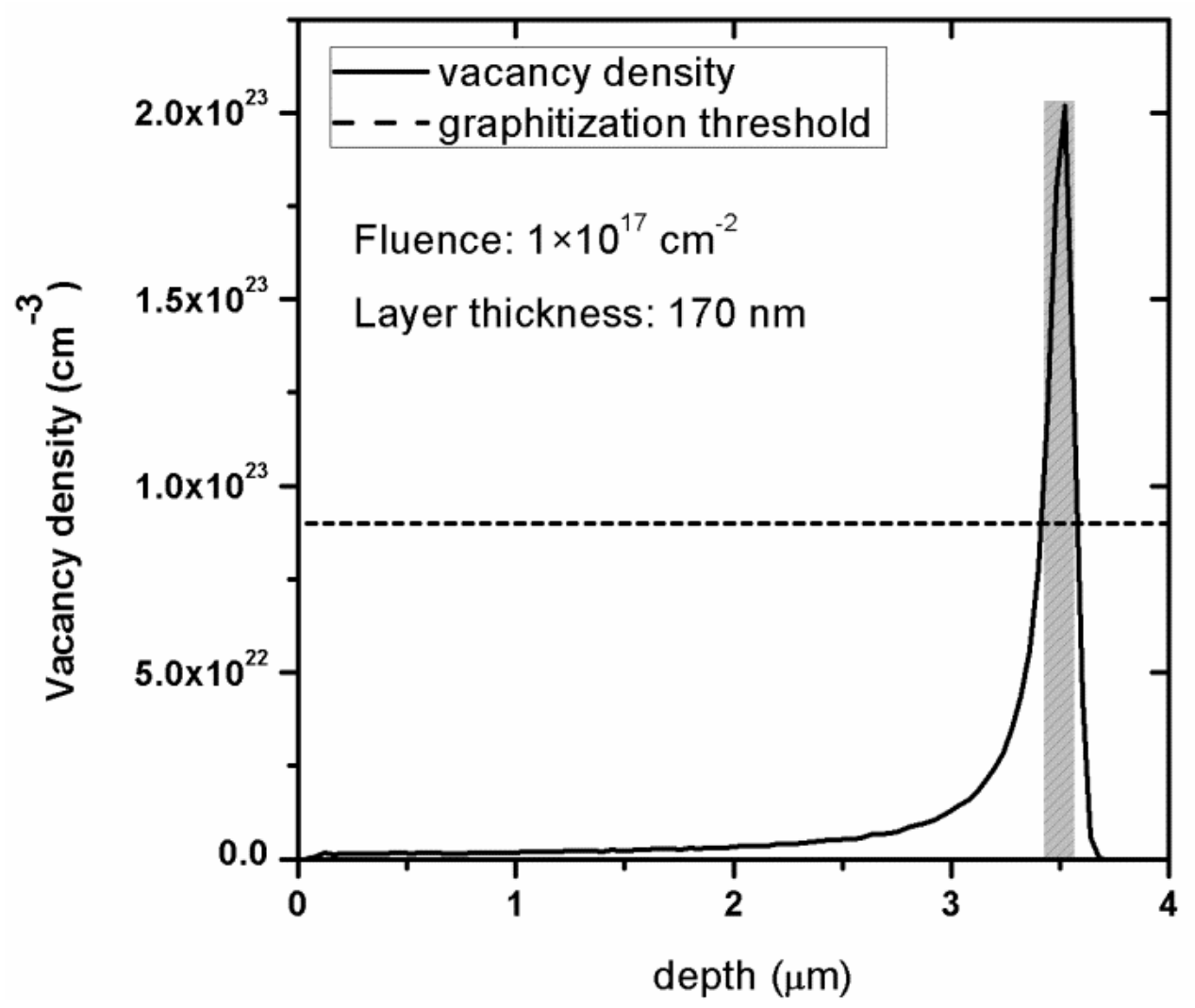

Fig. 1: Depth profile of the volumetric vacancy density induced in diamond by 2 MeV $He^+$ implanted at a fluence of $1\times10^{17}$ $cm^{-2}$. The graphitization threshold is reported in dashed line. The amorphized region is highlighted by the grey area in correspondence of the intersection of the Bragg peak with the graphitization threshold.

The above-mentioned implantation fluence allowed to overcome a critical damage density, usually referred as "graphitization threshold" [42], whose value in the above-mentioned linear approximation has been estimated as ~$9\times10^{22}$ $cm^{-3}$ for light MeV irradiation [43], as indicated in Figure 1. Such a model of the damage profile has to be considered as a rough estimation since it results from a linearly cumulative effect of ion damage, i.e. by neglecting any damage saturation effects occurring at high damage levels such as self-annealing and vacancies interactions [44,45]. Nonetheless, in this context it can be considered as a satisfactory approach to estimate the depth and thickness of the buried region. After ion implantation, the sample was thermally annealed for 1 hour at a temperature of 900 °C, which is suitable for the conversion of amorphous carbon to a graphitic phase, as confirmed by TEM studies [46–48]. Concurrently, the annealing process restores the pristine diamond structure in the lightly-damaged cap layer, i.e. the region comprised between the surface and the buried graphitic layer [49,50]. The process was carried out in vacuum ($p \approx 10^{-6}$ mbar) to avoid accidental etching of the diamond surface due to oxidation.

The ion-implanted side of the sample was subsequently irradiated with nanosecond-pulsed Nd:YAG laser (EzLaze3 by New Wave) equipped with a Q-switching system. This laser source generates pulses of 4 ns duration with a repetition rate of either 1 Hz or 5 Hz. Two different emission wavelengths can be selected, i.e. 1064 nm and 532 nm, the second one being obtained by means of an angle-tuned KTP crystal. The laser beam is focused onto the sample with a microscope supplied with 5×, 20× and 100× objective lenses. The co-axial imaging through the microscope offers the opportunity of monitoring the sample processing in real time. Using the 532 nm wavelength and the 100× objective, the maximum emission power is 150 Kw and the minimum size of the spot is $5\times5\ \mu m^2$, thus resulting in a maximum power density of ~22 GW $cm^{-2}$.

A Quanta 3D™ dual-beam system by FEI available at the "NanoFacility Piemonte" laboratories of the INRiM Institute was employed to cross-section the sample by 30 keV $Ga^+$ focused-ion-beam (FIB) milling and to estimate the thickness of the graphitic layer before and after the laser irradiation by SEM imaging .

The selective electrochemical etching of the graphite was performed with the purpose of enhancing the topographical contrast in the SEM imaging of the buried layers. This process was performed by immerging the sample in a water solution of $H_3BO_3$ ($4\times10^{-3}$ mol $l^{-1}$ concentration) for 1 hour applying a DC voltage of 150–200 V through a couple of platinum electrodes placed in close proximity of the sample [51].

The TEM imaging was performed at the Microscopy laboratories of the Bio21 Institute (University of Melbourne) for a detailed study of the thickness and the structure of the graphitic layers before and after laser annealing. To this scope, a Tecnai TF20 electron microscope operated at 200 keV was employed. Cross-sectional TEM samples with thickness ~100 nm were prepared in [110] and [100] orientations using a standard FIB lift-off technique. Selected area diffraction (SAD) patterns were collected with smallest aperture (diameter ~180 nm in the specimen plane). Nano-beam diffraction patterns were collected in nano-beam scanning TEM (STEM) mode with beam size ~10 nm. Electron energy loss spectroscopy (EELS) was conducted by employing a Gatan Enfina energy filter.

## 3. Results and discussion

FIB milling was employed to allow the cross-sectional SEM imaging of the ion-implanted and subsequently annealed regions, as shown in Fig. 2a. The presence of the graphitic layer embedded in the diamond matrix can be recognized in the dark grey horizontal strip, although the low contrast in the secondary emission yield between graphite and diamond does not allow performing an accurate measure of its thickness.

Selective etching of the graphite was therefore performed in order to improve imaging contrast and facilitate the measurement of the thickness of the sub-superficial layer. Fig. 2b shows a SEM micrograph of the same region of the sample reported in Fig. 2a, after selective chemical etching of the graphite layer. The previously graphitized region now corresponds to a gap within the material, thus allowing a better visibility of its thickness and depth.

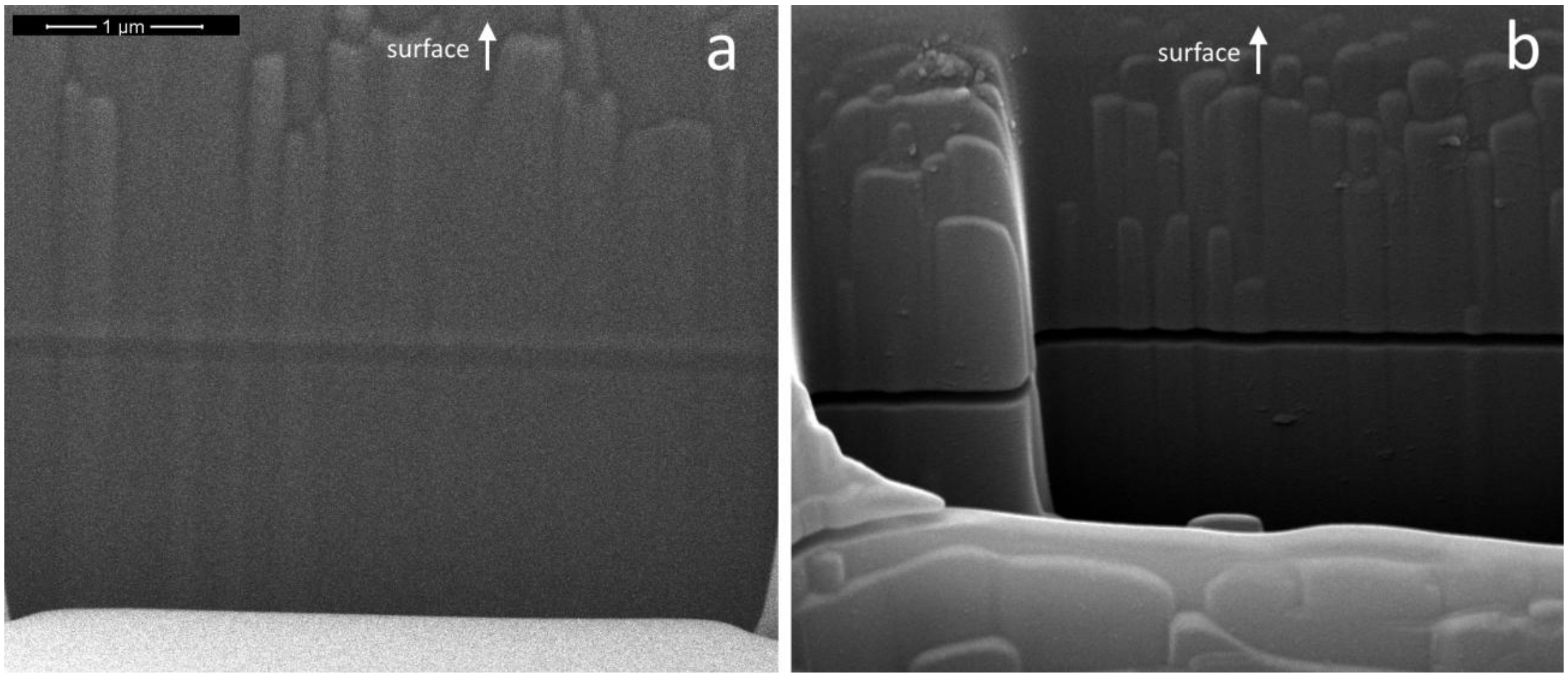

Fig. 2: Cross-sectional SEM micrographs of the sub-superficial graphitic layer in ion-implanted and annealed diamond a) before and b) after the selective etching process. A better definition of the geometrical parameters of the buried structure is clearly obtained after its selective removal by means of the etching process.

The thickness of the buried layer evaluated from Fig. 2a is (200 ± 40) nm, while it was estimated (140 ± 20) nm from Fig. 2b. This discrepancy is due to the relaxation of the upper cap layer, as it will be more extensively discussed in the following.

A more detailed structural characterization of the investigated regions was performed by means of cross-sectional TEM microscopy and EELS spectroscopy. A bright-field TEM micrograph and the corresponding Selected Area Diffraction (SAD) pattern of the implanted layer after thermal annealing are shown in Figs. 3a and 3b, respectively. The implanted layer is clearly visible due to the higher contrast and its width was estimated as (130 ± 3) nm. As shown in Fig. 3b, the diffraction pattern is characterized by well-defined spots arising from the diffraction of the bulk diamond, as well as arcs along the {220} reflections from the graphitic C-planes of the implanted layer. As shown in Fig. 3c, the corresponding EELS spectrum from the buried graphitic layer of the same region has a prominent feature at ~285 Ev, which is characteristic of $sp^2$ bonding. Therefore, TEM diffraction pattern combined with EELS confirm the conversion of the implanted layer into a nanocrystalline graphitic phase with predominant orientation of C-planes normal to the diamond surface.

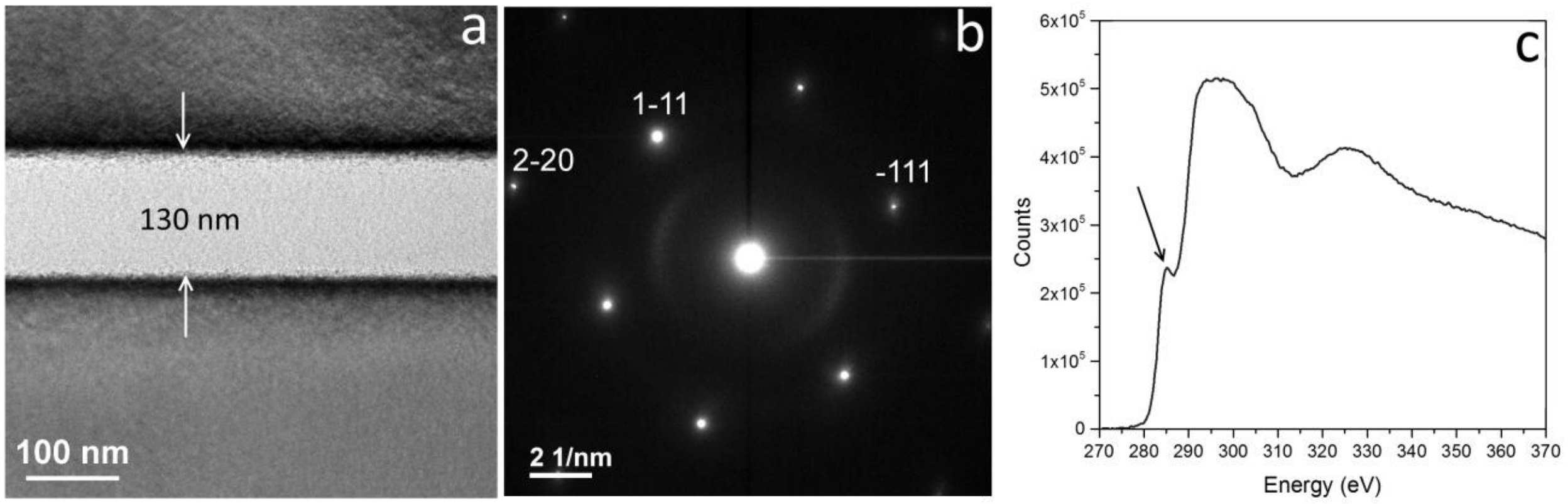

Fig. 3: a) Bright-field TEM cross-section micrograph of the sub-superficial graphitic layer in diamond before laser irradiation (the sample surface is located towards the top of the picture). The thickness of the ion-implanted and subsequently annealed structure is highlighted, corresponding to (130 ± 3) nm. B) SAD pattern taken from the same area, showing well defined spots and arcs respectively from the diamond and graphitic regions. C) EELS spectrum of the carbon K-edge taken from the implanted layer; the characteristic feature at 285 eV is indicated by the arrow.

Multiple-pulse laser irradiation at $\lambda$ = 532 nm wavelength of the ion-implanted and subsequently annealed diamond sample was carried out at two different powers densities (namely, 0.41 GW $cm^{-2}$ and 0.45 GW $cm^{-2}$) over an area of 26×26 $\mu m^2$.

Cross-sectional bright- and dark-field TEM micrographs of an area irradiated with 50 laser pulses at a power density of 0.41 GW $cm^{-2}$ are reported in Figs. 4a and 4b, respectively, and show a complex multi-layer structure within the buried layer. The laser-induced graphitic layer, formed in the region located directly above the ion-implanted layer, results in an overall thickness of (690 ± 15) nm. At the bottom of Fig. 4b (i.e. towards the bulk of the sample), the nanocrystalline graphite layer due to ion implantation is clearly distinguishable, exhibiting the same thickness measured before the laser irradiation (see Fig. 2) and no appreciable differences in its structure.

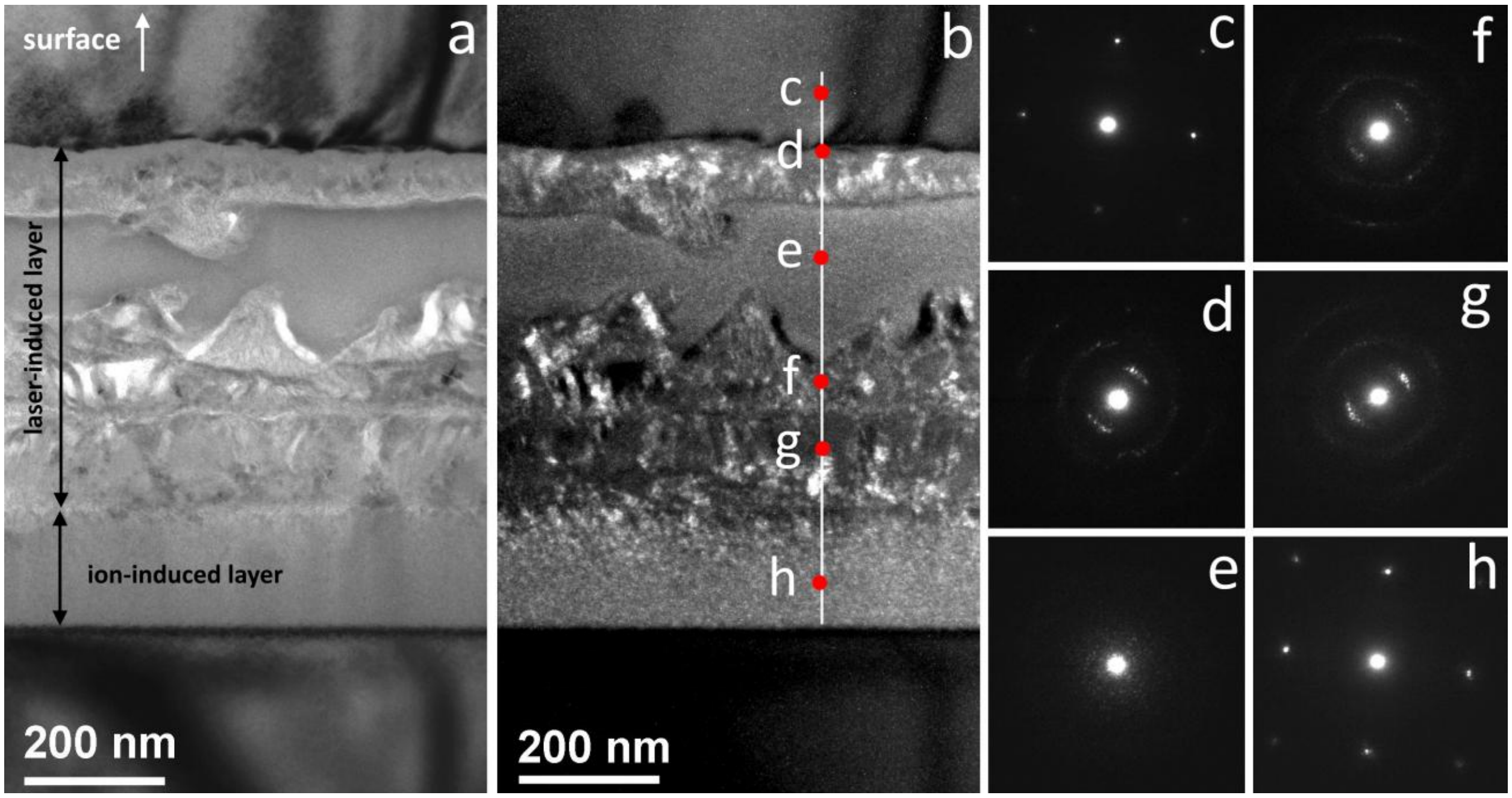

Fig. 4: a) Bright- and b) dark-field TEM cross section micrographs of a laser irradiated (λ = 532 nm, power density: 0.41 GW cm$^{-2}$, 50 pulses) nanocrystalline graphitic sub-superficial layer obtained after annealing at 900 °C of an implanted diamond region. The layer thickens towards the sample surface (i.e. towards the top of the figure), with a complex structuring of the modified region consisting in amorphous-carbon and nanocrystalline-graphite layers. The numbered red dots in b) highlight the regions from which the diffraction patterns reported in c-h) were acquired.

The laser-induced phase transition is driven by thermal effects that originate as a consequence of the strong absorption of the laser radiation by the implanted graphitic layer. Since the laser beam irradiates the side of the nanocrystalline graphitic layer facing the sample surface, the induced heating of the material is mainly localized within the "cap layer" comprised between the layer itself and the sample surface rather than towards the bulk, and therefore graphitization mainly occurs in such direction. An estimate of the temperature rise in the cap layer can be obtained from the pressure-temperature phase diagram of elemental carbon [52], with a similar approach to what is reported in [2]. Finite element simulation studies from diamond samples implanted under the same experimental conditions [53] indicate that the regions surrounding the buried graphitic layer experience pressures of 8–10 GPa due to its constrained volume expansion. Furthermore, the experimental evidence indicates that the temperature rise is high enough to drive the graphitization process within the corresponding stable portion of the phase diagram, without incurring in its liquefaction. The laser-induced local temperature rise is therefore estimated between 2500 °C and 4500 °C. Such a significant heating is expected to rapidly dissipate due to the high thermal conductivity of diamond.

Nano-beam diffraction patterns were taken in scanning TEM mode along the line shown in Fig. 4b, which is crossing perpendicularly the graphitic layer at 10 nm steps. A selection of the obtained diffraction patterns are shown in Figs. 4c-h. The peculiar structure of the laser-induced layer is characterized by the presence of an amorphous carbon phase (diffraction pattern in Fig. 4e) comprised between two polycrystalline graphitic phases (diffraction patterns in Figs. 4d and 4f), as clearly visible in the dark-field micrograph of Fig. 4b. The dark field micrograph was constructed by selecting the graphitic arcs in the diffraction pattern, and therefore the graphitic crystals appear as bright spots in it. These spots have <5 nm sizes in implanted layer and much larger dimensions in the laser-induced layer. The observed multi-layer structure is somewhat surprising and can be qualitatively attributed to a complex combination of thermal gradient and stress effects occurring in the diamond during the pulsed laser irradiation [54].

A SEM micrograph of the same laser-irradiated region is reported in Fig. 5a, from which a layer thickness of (700 ± 70) nm can be estimated, compatibly (although with lower accuracy) with what obtained from TEM imaging. The amorphous areas are visible inside the graphitic layer, as well as some voids. These voids were probably filled with re-deposited material during TEM sample preparation and are therefore not visible in TEM images reported in Figs. 4a and 4b.

Following the same procedure adopted for the sample after implantation and annealing (see Fig. 2), a selective electrochemical etching was carried out and the resulting structure is shown in the TEM micrograph reported in Fig. 5b. The removal of the graphitic layer is evident in both micrographs. Also, re-deposited material is visible on both surfaces of the gap, as commonly observed in FIB milling. As shown in Fig. 5b, after laser irradiation and chemical etching the buried layer was not entirely removed, but rather only the ion-induced nanocrystalline graphite layer was etched, as confirmed by the comparison of the thicknesses of the regions reported in Figs. 4a and 5b. This is a surprising result, since the polycrystalline graphitic phase formed during laser irradiation is expected to be effectively etched by the electrochemical attack.

A thermal treatment at 900 °C for 2 hours was therefore performed on the same sample after laser irradiation, with the purpose of inducing the graphitization of residual amorphous/distorted material in the un-etched region. Afterwards, the sample was exposed to the same etching process. As shown in Fig. 5c, the width of the etched layer in the laser-irradiated and annealed sample increased up to (420 ± 20) nm, indicating a full removal of the laser-irradiated layer. As already reported for the sample before laser irradiation, a discrepancy between the thickness values of the laser-induced layer and the remaining gap is observed (see Figs. 5a and 5c). As mentioned before, this effect can be explained by considering that a relaxation of the diamond cap layer takes place after the removal of the graphitic layer. Since the graphite is characterized by a significantly lower atomic density with respect to diamond, a volume expansion takes place upon the graphitization process [55,56], thus deforming the upper diamond cap layer comprised between the graphitic layer and the sample surface. Once the graphite is removed, the cap layer undergoes a structural relaxation which induces the thinning of the etched region.

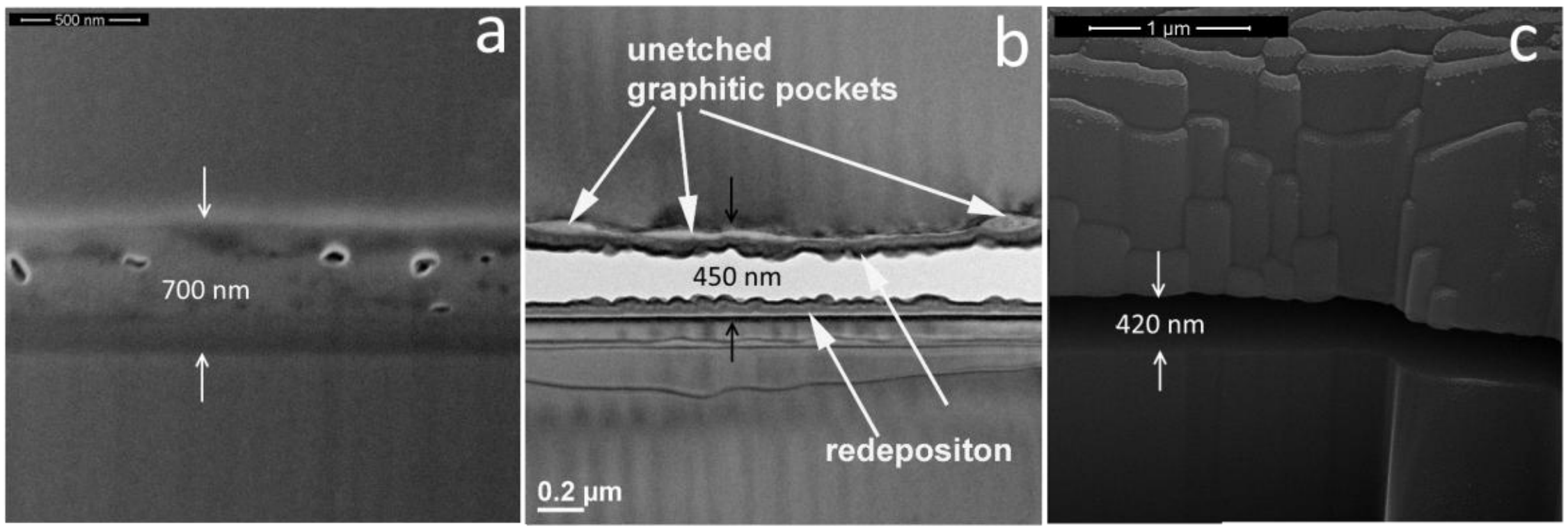

Fig. 5: a) Cross sectional SEM micrograph of the buried damaged layer after laser irradiation (λ = 532 nm, power density: 0.41 GW $cm^{-2}$, 50 pulses). b) Cross-sectional TEM micrograph of the same layer reported in a) after electrochemical etching; together with the characteristic re-deposited material present in the gap, un-etched laser-induced layers are recognizable around the gap. c) Cross-sectional SEM micrograph of the same layer reported in a) after a further 900 °C annealing step and electrochemical etching; no un-etched layers are visible.

Even more surprisingly, after the first etching process (i.e. before the second annealing step) highly oriented graphitic clusters were found in correspondence of the partially un-etched diamond/graphite interface which was closest to the sample surface, as shown in Fig. 5b.

Fig. 6a reports a high-resolution TEM micrograph of one of these clusters. The presence of graphite C-planes is clearly visible in the corresponding diffraction pattern reported in Fig. 6c, indicating the highly oriented structure of these graphitic clusters.

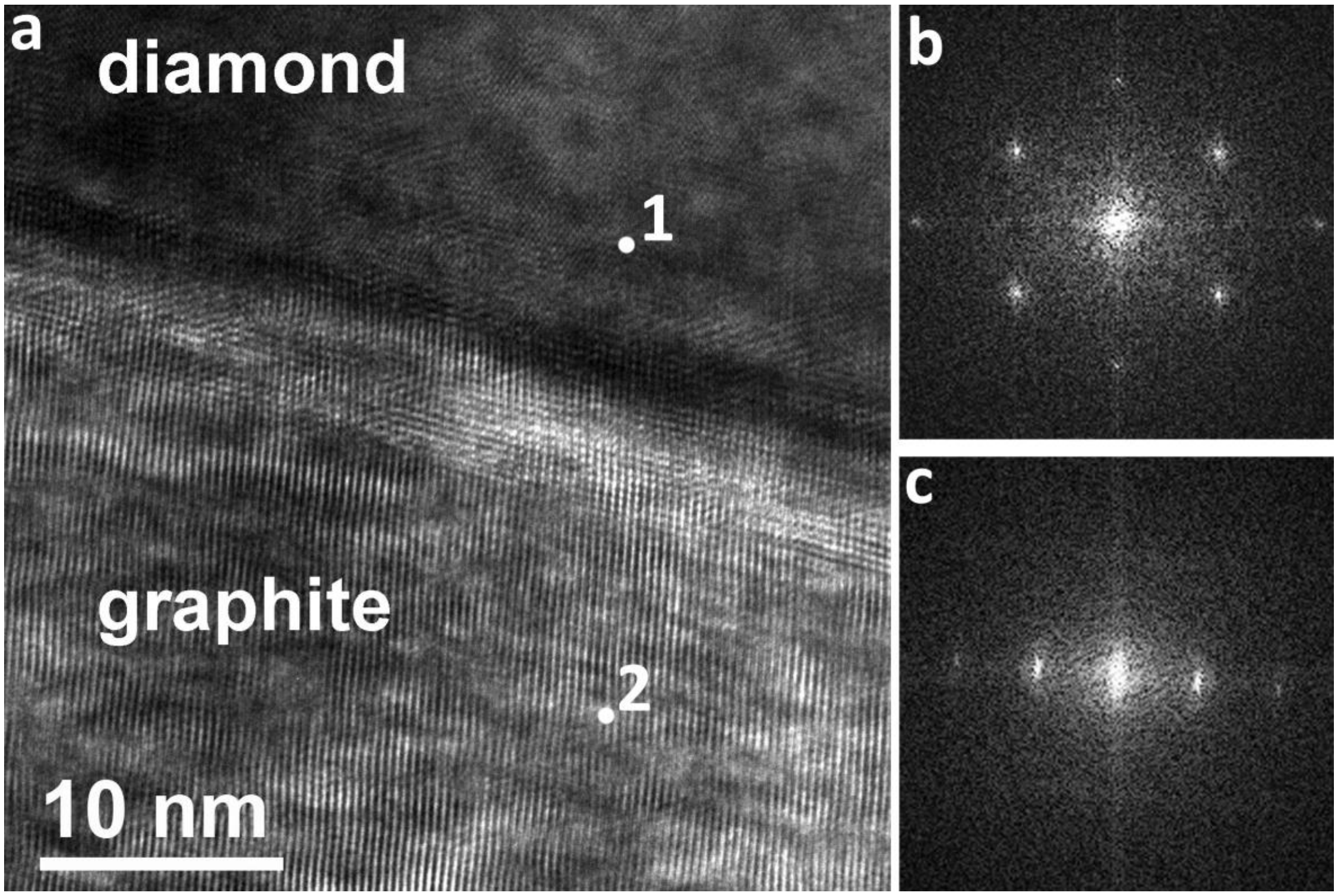

Fig. 6. a) High resolution TEM micrograph of one of the highly oriented graphitic clusters remaining near the diamond interface after the chemical etching of the laser-irradiated sample. The corresponding diffraction patterns from diamond (spot "1") and graphite (spot "2") are reported in b) and c), respectively.

The results reported so far refer to regions of the sample which were exposed to 50 laser pulses and subsequently subjected to different annealing and/or etching processes. A systematic investigation of the laser-induced graphitization was also carried out as a function of the laser irradiation parameters (power, number of pulses). As previously reported, the implanted and subsequently annealed sample was irradiated with 532 nm laser pulses and power densities of 0.41 GW $cm^{-2}$ and 0.45 GW $cm^{-2}$ over different 26×26 $\mu m^2$ areas. Following these laser irradiations, the sample was annealed at 900 °C for 2 hours with the purpose of inducing a full graphitization in the irradiated areas, consistently with the above-mentioned results. Also in this case, cross-sectional SEM imaging of the irradiated regions was systematically performed to evaluate the thickening in regions processed with different laser irradiation conditions, and the layer thicknesses were measured before and after the etching process.

Fig. 7 summarizes the results obtained for the two different power densities for a number of laser pulses ranging between 1 and 300. It is evident that the thickness of the graphitic sub-superficial layer obtained after laser irradiation and thermal annealing increases at increasing numbers of laser pulses, up to more than 6 times its initial value. The trends reported in Fig. 7 clearly indicate a strong sub-linear dependence of the layer thickness from the number of laser pulses. It is also evident that the power density has a direct influence on the thickening process, with the larger power density determining a more pronounced thickening for the same number of laser pulses. A maximum total thickness of ~1.25 μm was achieved, with the process being limited by the large mechanical stresses that build up in the diamond cap layer which ultimately cause local mechanical fractures. For this reason, power densities and number of pulses larger than the ones shown in Fig. 7 could not be tested without incurring in structural damage effects. Consistently with what was previously observed, for all of the structures the second annealing step resulted in a complete removal of the buried graphitic layers upon electrochemical etching. It is also worth stressing that, as explained above, the thickness of the gaps obtained after selective electrochemical etching is systematically smaller than the thickness of the laser-induced graphitic layers, as shown in Fig 7.

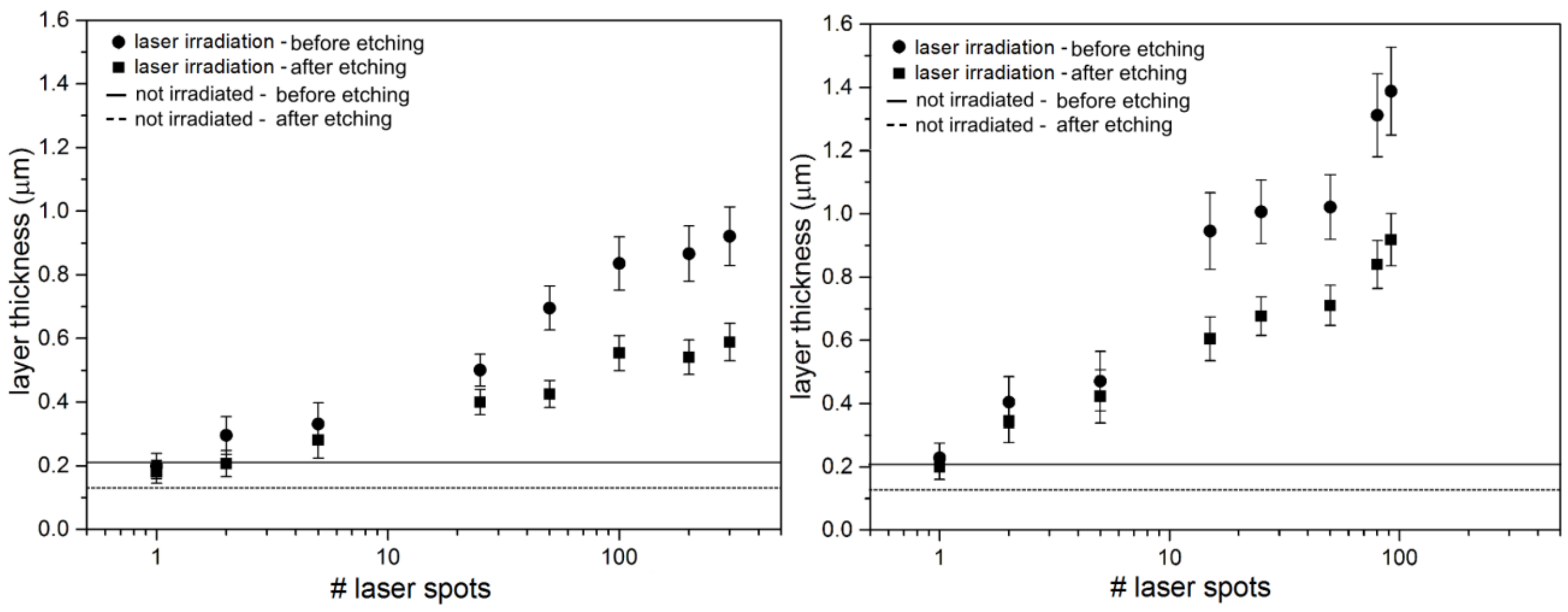

Fig. 7: Plot of the measured thickness of the sub-superficial graphitic layers versus the number of λ = 532 nm laser pulses for two different power densities, i.e. a) P = 0.41 GW $cm^{-2}$ and b) P = 0.45 GW $cm^{-2}$. The thickness measurements are reported both before (circular dots) and after (square dots) the selective etching process. Likewise, the horizontal lines indicate the thicknesses of the pristine ion-induced layer, both before (continuous line) and after (dashed line) the selective etching process.

## 4. Conclusions

We reported on the effect of ns-pulsed λ = 532 nm laser irradiation on the thickening of sub-superficial graphitic layers in diamond obtained by means of 2 MeV $He^+$ ion implantation and subsequent high-temperature annealing. Cross-sectional TEM and EELS measurements elucidated the complex structuring of the processed regions into amorphous and nanocrystalline graphitic multi-layers. The complete conversion of the laser-induced layers to an etchable graphitic phase was obtained only upon thermal annealing at 900 °C, while a highly-oriented phase was found in the residual graphitic pockets after selective etching of the non-annealed samples. A systematic SEM investigation of the thickening of the graphitic region was carried out as a function of the laser irradiation parameters (power density, number of pulses). An increase up to 650% of the initial layer thickness was reached without incurring into critical mechanical failures due to induced mechanical stresses.

By allowing a fine tuning of geometrical and structural properties of graphitic layers formed by ion irradiation, laser-induced graphitization offers interesting opportunities for a new level of control in the fabrication of buried graphitic structures in diamond, with appealing applications in different fields in which MeV ion beam lithography and laser graphitization were successfully employed [26,31–36,38,39]

**Acknowledgments**

The authors wish to thank Genny Giaccardi for the kind support during laser processing. This work is supported by the following projects: "DiNaMo" (young researcher grant, project n° 157660) by National Institute of Nuclear Physics (I); FIRB "Futuro in Ricerca 2010" (CUP code: D11J11000450001) funded by MIUR and "A.Di.N-Tech." (CUP code: D15E13000130003) project funded by the University of Torino and "Compagnia di San Paolo". The Nanofacility Piemonte laboratory is supported by the "Compagnia di San Paolo" foundation.